# Fluid-like Surface Layer and Its Dynamics in Glassy Nanotubes

*Matthew C. Wingert[a], Soonshin Kwon[a], Shengqiang Cai[a,*], and Renkun Chen[a,*]*

[a] Department of Mechanical and Aerospace Engineering, University of California, San Diego, La Jolla, California 92093, United States

[*]E-mail: shqcai@ucsd.edu; rkchen@ucsd.edu

**Abstract**

We observed strongly size-dependent viscoelasticity in amorphous $SiO_2$ and Si nanotubes with shell thickness down to ~8 nm. A core-shell model shows that a ~1 nm thick fluid-like surface layer has a significant effect on the mechanical behavior of nanotubes and matches well with our experimental results. Surprising, the surface layer exhibits a room temperature viscosity equivalent to that of bulk glass above 1000 °C. Additionally, a low activation energy extracted from temperature dependent creep tests indicates that the viscous flow in the surface layer is due to bond motion/switching, instead of bond breaking. These findings unambiguously show the presence of a fluid-like surface layer and elucidate its role on dynamic mechanical behavior in nanoscale inorganic glass.

Nanoscale glassy films are ubiquitous and critical components for a variety of electronic and energy devices such as optical coatings[1-3], amorphous oxide transistors[4, 5] and gate dielectrics[6-8], non-volatile solid state memory[9, 10], memristors[11-13], solid-electrolyte interphase layers[14] and anodes[15, 16] in batteries, and amorphous Si solar cells[17, 18]. The mechanical properties of such amorphous materials are practically important in system design and issues can arise when shrinking them down to nanoscale dimensions. For instance, in glassy polymers, extensive studies have shown the presence of a highly mobile surface layer on the order of tens of nanometers thick.[19-26] In the case of glassy (or amorphous) inorganic materials, whether or not a similar surface layer exists, however, is largely unexplored. A prior study has indicated such a possibility through imaging the motion of surface atoms in ultrathin $SiO_2$ films[27, 28], but its effect on the dynamic mechanical properties is not known. On the other hand, it is well known that inorganic nanostructures, in both amorphous and crystalline forms, can exhibit unusual mechanical phenomena, such as electron beam induced deformation[29-31], brittle-plastic transition[32-34], and strong anelasticity (atomic diffusion)[35]. It is, therefore, expected that if a mobile surface layer does exist, the mechanical behavior of nanostructured glass could be different from its bulk form. One of the challenges of directly probing the dynamic mechanical response of this surface layer lies in creating and measuring ultra-thin samples, as the mobile surface layer, if it exists, would be on the scale of several nanometers, much smaller than that in polymers.

Herein, we systematically studied the dynamic mechanical behavior of novel amorphous $SiO_2$ (a-$SiO_2$) and Si (a-Si) nanotube (NT) structures with exceedingly thin shells down to 7 nm (Figure 1), unambiguously showing the presence of highly mobile surface layers in inorganic glasses. The thin shells and high aspect-ratio of the tubular structure, having increased surface atom fraction, allow us to apply large stresses ($10^9$ Pa) and examine the mechanical behavior. Exponentially increasing creep deformation at room temperature is observed as the NT shell thickness is reduced, with strain increases up to 60% greater than the initial strain (< 5.5%) for the thinnest tubes. We also observed that creep deformation occurs only in the amorphous samples we studied (a-SiO$_2$ and a-Si NTs) and not in crystalline Si (c-Si) NTs of similar geometry. Furthermore, we performed temperature dependent creep tests and found a low diffusion-like activation energy associated with the creep.

The ultrathin NTs in this study were fabricated by etching the Ge cores in Ge-Si core-shell nanowires (NWs)[36] as we previously reported, where crystalline and amorphous NTs were fabricated by controlling the temperature during epitaxial deposition of the Si shells on the Ge cores.[37] Amorphous SiO$_2$ NTs were synthesized by oxidizing c-Si NTs in air at 950 °C for 3 hours and are completely oxidized with no internal interfaces while the a-Si and c-Si NTs have 1-2 nm thick inner and outer native oxide coatings encompassing the Si shells. NT shell thicknesses ranged from ~7.5, 12, and 20 nm for a-SiO$_2$, ~9, 12, 20, and 150 nm for a-Si, and ~10 nm for c-Si, ±2 nm each, with deviations of < 1 nm along the NT length. The outer diameters (ODs) of the NTs range from ~70-150 nm. Each NT was characterized by high-resolution (HR) TEM (representative images in Figure 1a, b, and c, with larger NTs shown in Supplemental Information[38]) prior to tensile testing in order to confirm the crystallinity of the sample and to precisely measure its dimensions (OD, shell thickness, and native oxide thickness for the a-Si and c-Si NTs).

After TEM characterization, the *same* exact NTs were then transferred to a dual (electron and ion) beam chamber where a micromanipulator probe was used to pick up and attach each NT to the free end of an AFM cantilever. Tensile measurements were then performed according to an established procedure[37, 39] using the setup shown in Figure 1d. During the initial loading stress-strain measurements were performed to calculate instantaneous tensile moduli[38], which were consistent with those of $SiO_2$,[40-42] a-Si,[43] and c-Si nanostructures,[44] including our previously measured NTs[37]. Each of these initial measurements only took 2-3 minutes and was performed in the linear-elastic regime, with strains and stresses up to 1.5-5.5% and 0.8-5 GPa, respectively.[38] Such large stresses can only be sustained when the nanostructure has a low volumetric concentration of defects to avoid fracturing, as shown by the high fracture strength of our NTs[38]. Transient creep was then monitored with the NT at the maximum loaded stress by measuring changes in length every 5 to 10 minutes over a period of up to two hours, blanking the electron beam between measurements. Several samples had gauge marks deposited *via* e-beam induced Pt (seen in Figure 1d) and strain was measured using both the gauge marks and the entire NT length to ensure no slippage occurred. Temperature dependent creep tests were also performed *in-situ* in the same chamber, where a heater implanted in the sample stage was used to control the temperature above 300 K and the sample temperature was measured with a thermometer.

The creep strain measured in the a-$SiO_2$ NTs is shown in Figure 2a, where the strain increase due to creep, $\Delta\varepsilon$, is normalized by the initial strain, $\varepsilon_o$. The majority of the creep occurred within the first 20-30 minutes and saturated thereafter. The creep is clearly related to the a-$SiO_2$ NT shell thickness ($\delta$), where $\Delta\varepsilon/\varepsilon_o$ only reaches 10% for NTs with $\delta$>10 nm but is as high as 60% when $\delta$<10 nm. The a-Si NTs exhibit a similar amount of creep in terms of overall strain increase and size-dependence, shown in Figure 2b. This similarity between the ultrathin a-$SiO_2$ and a-Si NTs indicates that a-Si indeed exhibits glass-like behavior, as has been previously suggested.[45, 46] The creep saturation in the a-Si NTs, however, took longer than in the a-$SiO_2$ NTs. Negligible creep was observed in a 145 nm thick a-Si NT, as expected for bulk-like behavior. Interestingly, no creep was observed in ~10 nm thick crystalline Si NTs (Figure 2c), just like in our control experiment on a 50 nm diameter c-Si NW. The viscoelastic response of the 7.5 nm a-SiO$_2$ NTs was studied by continuously measuring the strain after the applied stress was un-loaded following creep saturation, with a representative measurement shown in Figure 2d. Most of the creep strain was recovered over the next 30 minutes, indicating that the creep deformation in this study is viscoelastic, as opposed to previously observed[34] enhanced plastic deformation which occurred above a threshold yield-stress and with larger strain (typically ~10%).

The size-dependent creep observed in Figure 2a, b, and d is presumably due to the growing influence of a fluid-like surface layer on the amorphous nanostructures, similar to that of the thin mobile surface layers in amorphous polymer films. As such, we used a core-shell model, similar to a Maxwell-solid spring-dashpot system (Figure 3a), to understand the observed behavior, where the NT shell is comprised of mobile surfaces surrounding a stiff center layer. Here, the time dependent mechanical response of the NT under the influence of a tensile force, $F(t)$, is

$$F(t) = E_1 \varepsilon(t) A_{core} + E_2 \varepsilon'(t) A_{surf} \qquad (1)$$

where the stiff middle layer has a modulus, $E_1$, cross-sectional area, $A_{core}$, and experiences a strain, $\varepsilon(t)$, while the fluid-like surface layers have elastic moduli, $E_2$, total cross-sectional area, $A_{surf}$, and experience a strain, $\varepsilon'(t)$. The average stress over the entire cross-section of the NT, with a total shell thickness of $\delta$, and inner/outer surface layers, each of thickness $c$, is

$$\bar{\sigma}(t) = E_1 \varepsilon(t)\left(1 - \frac{2c}{\delta}\right) + E_2 \varepsilon'(t)\left(\frac{2c}{\delta}\right) \qquad (2)$$

and the elastic and viscous contributions in the inner/outer surface layers are related by

$$E_2 \varepsilon'(t) = \eta \dot{\varepsilon}''(t) \tag{3}$$

where $\eta$ is the viscosity of the surface layer and $\varepsilon''(t)$ is the viscous strain. Using Eqs. 2 and 3 and the condition of constant average stress, the time-dependent creep strain in the NT, normalized by the initial strain, can be found to be

$$\frac{\varepsilon(t)}{\varepsilon_o} = 1 + \frac{E_2\left(\frac{2c}{\delta}\right)}{E_1\left(1-\frac{2c}{\delta}\right)} - \left[\frac{E_2\left(\frac{2c}{\delta}\right)}{E_1\left(1-\frac{2c}{\delta}\right)}\right] \cdot \exp\left[-\frac{E_1\left(1-\frac{2c}{\delta}\right)}{E_1\left(1-\frac{2c}{\delta}\right)+E_2\left(\frac{2c}{\delta}\right)}\left(\frac{t}{\tau}\right)\right] \tag{4}$$

where the time constant is defined as $\tau=\eta/E_2$ and the normalized maximum creep strain is

$$\frac{\Delta \varepsilon_{max}}{\varepsilon_o} = \frac{E_2}{E_1}\left(\frac{2c}{\delta - 2c}\right) \tag{5}$$

We assumed that E1≈E2 based on the equation 2 at t=0 ($\varepsilon = \varepsilon'$) and size-independent instantaneous modulus[38] (Figure S4A) measured from the amorphous NTs and found that a surface layer thickness (*c*) of 1.1 nm best fits the experimental data for both amorphous NT materials, which is reasonable considering that the mobile surface layers should be on the order of nanometers. The creep calculated from Eq. 4, and represented by the dashed lines in Figure 2a and b, matches the measured time-dependence trends of the creep fairly well. The initial creep shown in Figure 2d also fits well to Eq. 4, where the initial starting strain was calculated as the ratio of the initial stress and elastic modulus, $\varepsilon_o=\sigma_o/E_1$. The creep recovery after stress removal is simply

$$\varepsilon_{recovery}(t) = \Delta\varepsilon_{max} \cdot \exp\left[-\frac{E_1\left(1-\frac{2c}{\delta}\right)}{E_1\left(1-\frac{2c}{\delta}\right)+E_2\left(\frac{2c}{\delta}\right)}\left(\frac{t-t_c}{\tau}\right)\right] \tag{6}$$

where $\Delta\varepsilon_{max}$ is defined from Eq. 5 and $t_c$ is the time at which the stress was removed. Here, the time constant that best fit the a-SiO$_2$ and a-Si NTs was 6 and 18 minutes, respectively. The corresponding viscosity of the soft outer/inner layers can be estimated from $\eta=\tau\cdot E_2$ and was ≈10-20 TPa·s for the a-SiO$_2$ NTs at room temperature, equivalent to the viscosity of glass measured[47,48] between 1250-1500 K, and ≈ 60-90 TPa·s for the a-Si NTs. This is an astonishingly low viscosity at room temperature for inorganic glass, given the exponential dependence of viscosity on temperature. One would not have predicted *a priori* the order of magnitude of room temperature viscosity of these surface layers in a-SiO$_2$ and a-Si, given the fact that the highly mobile layer in glassy polymers occurred at a temperature (room temperature) only a few tens of degree K below their glass transition temperature.

The size-dependence of the creep deformation is more clearly evident in Figure 3b, where $\Delta\varepsilon_{max}/\varepsilon_o$ is plotted against NT shell thickness. The a-SiO$_2$ and a-Si NTs follow strikingly similar size-dependent trajectories, with the maximum creep largely increasing as the NT shell thickness decreases. The observed creep trends only with the NT thickness and is independent of testing parameters such as the initial strain and stress.[38] The dashed lines represent the modeled saturation creep strain based on Eq. 5 using the same constants previously mentioned and shows good agreement with the experimental data. Interestingly, if we calculate the *relaxed* tensile modulus using the saturated total strain,[38] we find that it decreases as $\delta$ falls below 30 nm for a-Si, similar to the size-dependent moduli of c-Si NTs[37] and NWs[39].

Differences in the saturation time between the a-SiO$_2$ and a-Si NTs, however, may stem from a difference in atomic species mobility in either material. It is also possible that the a-SiO$_2$/a-Si interface, present in the a-Si NTs and a source of atomic defects and roughness[37], may frustrate overall atomic mobility, increasing the time required for the atoms to reorganize.

Furthermore, the lack of any creep in the c-Si NTs, even with similar geometry as the a-Si NTs and with native surface oxides, also highlights the importance of the exact nature of this interface. This could be caused by the presence of a higher density layer at the c-Si-SiO$_2$ interface (~1nm thick)[49], or atoms in the native surface oxide, near to the interface with the ordered crystalline surface, could be more ordered, confining and inhibiting their movement.[50] However, understanding the exact role of the interfaces among the a-SiO2, a-Si and c-Si NTs warrants further investigation.

To understand the mechanism behind the surface-based creep, we performed temperature dependent creep tests on 20 nm thick SiO$_2$ NTs (Figure 4a), and found that the amount of creep significantly increased with temperature, however $\tau$ remained constant. The resulting creep viscosity was found to fit well to an Arrhenius temperature dependence[51] (Figure 4b),

$$\eta = E_2\tau \propto \exp\left(\frac{E_A}{k_B T}\right) \qquad (7)$$

where $k_B$ is the Boltzmann constant and $T$ is absolute temperature. The activation energy of the deformation processes, $E_A$, was found to be $0.1 \pm 0.05$ eV, of similar order as low values for diffusion and defect relaxation (~0.2 eV) in a-Si[52-54] and a-SiO$_2$[55]. Furthermore, activation energies could be lower for the fluid-like surface, similar to values simulated for SiO$_2$ surfaces[56,57]. These energies are consistent with analysis[58] showing deformation at room temperature requires activation energies less than 0.3 eV and indicate that bond motion, not bond breaking[59], is the source of the measured creep.

Based on our low observed activation energy the creep deformation in the amorphous NTs is most likely due to multiple defected bond relaxation events which result in the migration of atomic clusters[60, 61]. Such reorganizations will alter coordinated bonding groups, rings of Si-O molecules in a-SiO$_2$ or Si-Si in a-Si, such as transforming five- and six- fold coordinated group rings into three-, four-, and eight- member coordinated groups[62] and migration of non-bridging atoms leading changes in bond angles[60]. This atomic re-ordering is particularly concentrated at surfaces, where natural defects such as dangling bonds allow atomic processes to circumvent the higher energy barriers present among the bulk atoms, leading to the high mobility of the surface layers. Since our NTs are made up of a significant fraction of surface atoms, they will increasingly govern the mechanical response of the entire nanostructure, as observed in in our experiments.

Deformation in prior studies on SiO$_2$ and Si nanostructures was investigated using constant strain rates[33, 34] or high energy electron beams[29-31] which break atomic bonds and dislocate atoms. One such recent study[34] found that sub-20 nm diameter a-SiO$_2$ nanofibers show ductile behavior and enhanced plastic elongation above a threshold yield-stress. Our results, however, were performed in the elastic range[38], and while we show similar strong size-dependent deformation, it occurs significantly below the yield-stress and is viscoelastic, recovering the majority of the creep upon removal of the tensile load. Also in this study, the 5 kV electron beam will only transfer <0.4-0.7 eV to atoms our nanostructures[63], which is much lower than the energy required to break Si-O or Si-Si bonds (~2-5 eV)[64], but could have some influence on our measurement due to our measured $E_A$ = 0.1 ± 0.05 eV. TEM irradiation in previous studies is much stronger, 100-300 kV beams can transfer ~9-33 eV,[63] strong enough to continuously create defects. We took further precaution by blanking the beam between measurements. Heating of the nanostructure should be negligible under a periodically scanning 5 kV beam. Stresses on the order of $10^9$ Pa, as applied in our experiments[38], are also large enough to initiate defect bond rearrangements, providing >0.4 eV compared to our 0.1 ± 0.05 eV.

We observed strongly size-dependent creep deformation in ultrathin amorphous nanostructures, which was viscoelastic and not solely plastic in nature, nor limited by a threshold yield-stress. This same time- and size- dependent behavior did not occur in crystalline structures with similar geometry and native surface oxide. Furthermore, a core-shell model shows that a ~1 nm thick fluid-like surface layer has a significant effect on mechanical behavior and matches well with our experimental results. Additionally, a low activation energy extracted from temperature dependent creep measurements indicates that the deformation is due to bond motion, not bond breaking. These findings show that the mobility of surfaces must be accounted for in nanoscale amorphous systems. Such behavior could have major implications for the design and functionalities of future electronic and energy devices.

**Acknowledgment:**

This work was supported by National Science Foundation (DMR- 1508420 for synthesis and experiments, CMMI-1538137 for modeling). FIB work was performed in Nano3 cleanroom at UCSD, a CALIT-2 facility. We thank FEI Company and Dr. B. Fruhberger and R. Anderson of Nano3 for the support and assistance on the high-temperature sample stage in FIB. M. C. W and S. K contributed equally to this work.

**Figure Captions**

Figure 1: TEM images of individual ultrathin (a) amorphous $SiO_2$, (b) amorphous Si, and (c) crystalline Si nanotube (NT) samples and (d) SEM of tensile creep experimental setup. Scale bars are (a) 10 nm, (b) 50 nm, and (c) 5 nm.

Figure 2: Creep strain measured for (a) amorphous $SiO_2$ (a-$SiO_2$), (b) amorphous Si, and (c) crystalline Si nanotubes (NTs), and (d) viscoelastic behavior of a 7.5 nm shell a-$SiO_2$ NT. Colored bands in (a) and (b) represent the range of the experimental data for similar shell thickness NTs. The insets of (d) show schematics of the loading and unloading of the NT with the AFM cantilever. The dashed lines in (a), (b), and (d) represent fittings based on the core-shell model shown in Figure 3a.

Figure 3: (a) Core-shell model describing the dynamic and size-dependent behavior of the amorphous nanotubes and (b) the experimental (circles and squares) and modeled (dashed line) size-dependent normalized saturated creep.

Figure 4: (a) Creep strain observed for 20 nm shell amorphous $SiO_2$ nanotubes at various temperatures. (b) Creep viscosity versus temperature, fitted to an Arrhenius temperature dependence with an activation energy of $E_A = 0.1 \pm 0.05$ eV, consistent with bond motion/rearrangement. For comparison, temperature dependent viscosity requiring an activation energy of 1 eV, signifying a bond breaking mechanism, is shown (short dashed line).

**Figures**:

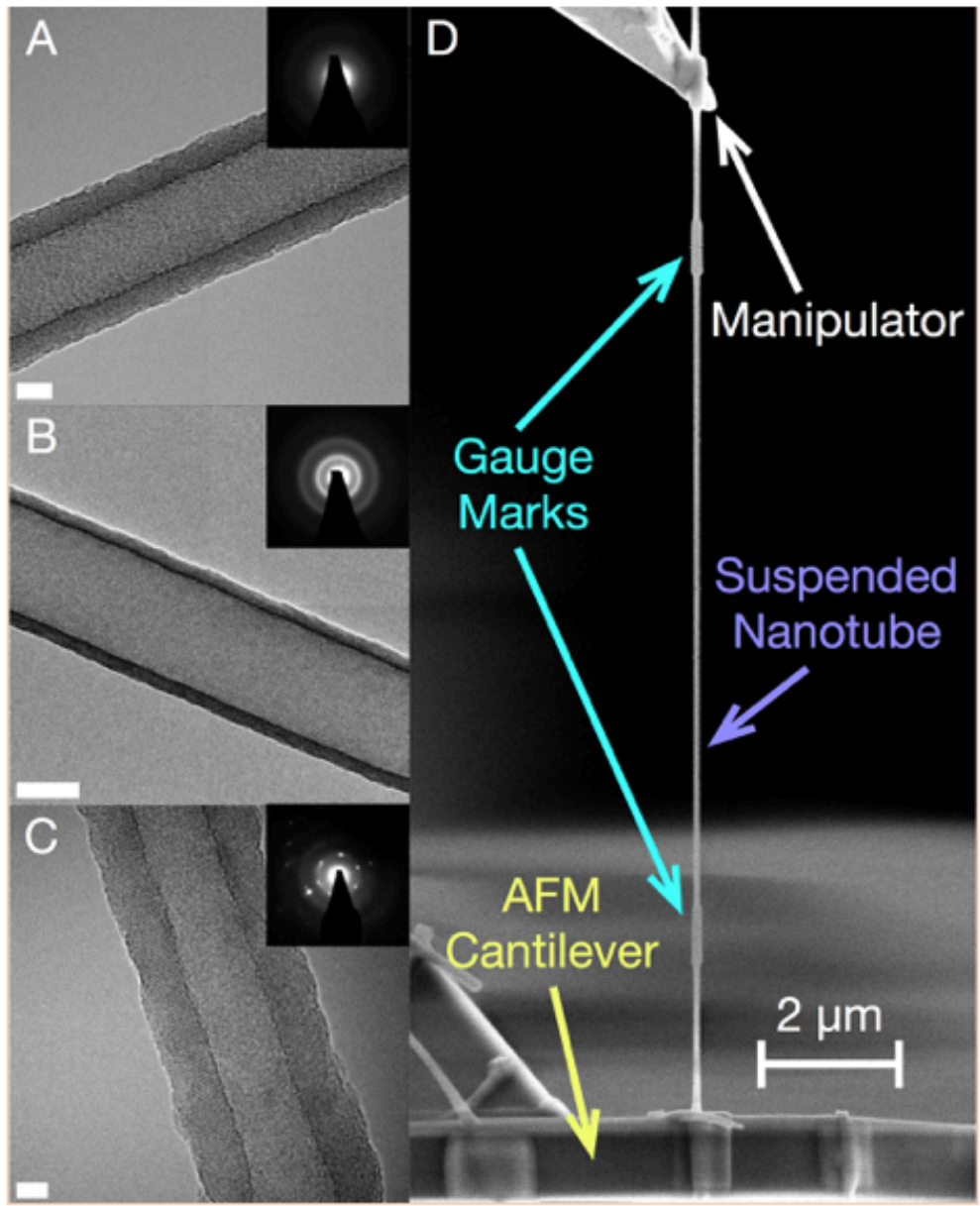

Figure 1: TEM images of individual ultrathin (a) amorphous $SiO_2$, (b) amorphous Si, and (c) crystalline Si nanotube (NT) samples and (d) SEM of tensile creep experimental setup. Scale bars are (a) 10 nm, (b) 50 nm, and (c) 5 nm.

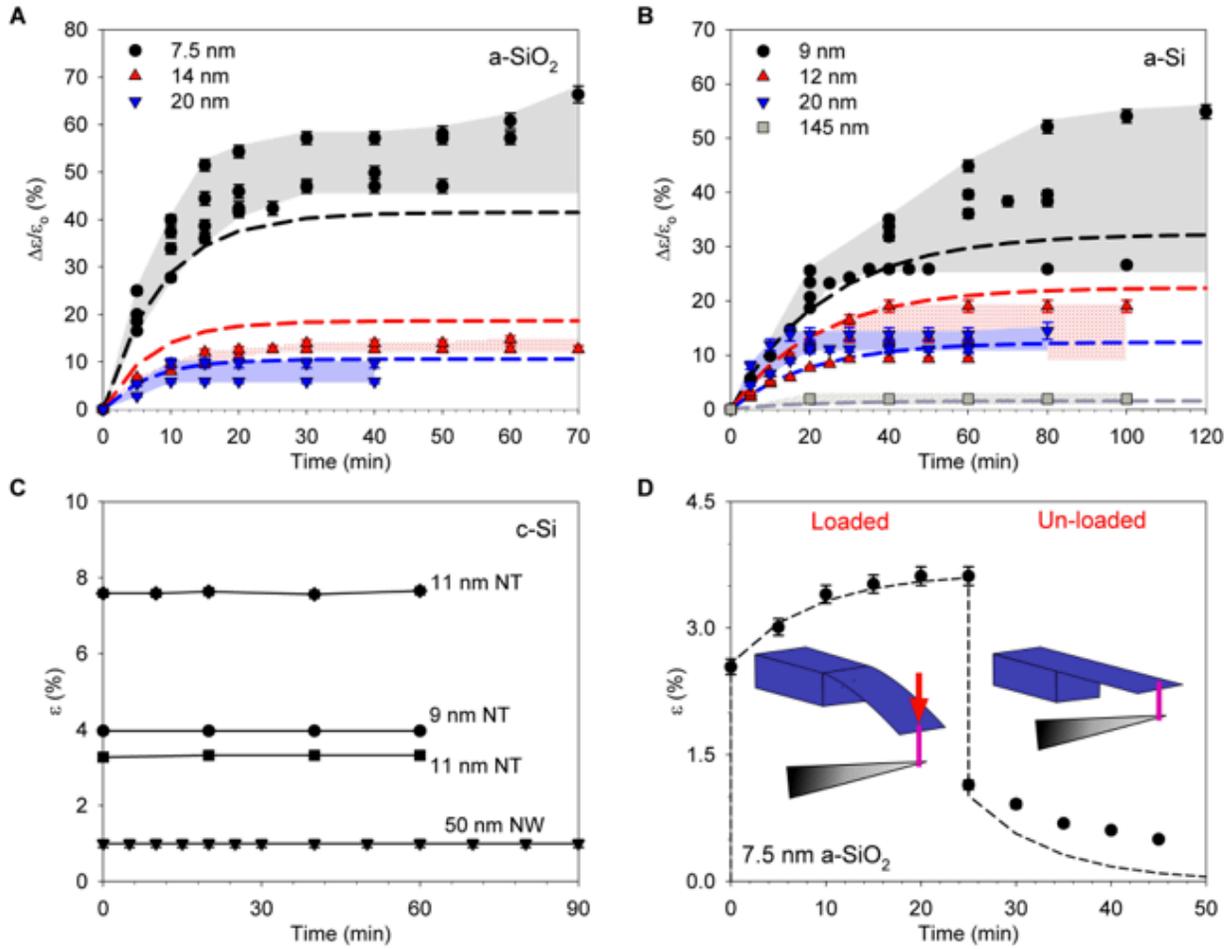

Figure 2: Creep strain measured for (a) amorphous SiO$_2$ (a-SiO$_2$), (b) amorphous Si, and (c) crystalline Si nanotubes (NTs), and (d) viscoelastic behavior of a 7.5 nm shell a-SiO$_2$ NT. Colored bands in (a) and (b) represent the range of the experimental data for similar shell thickness NTs. The insets of (d) show schematics of the loading and unloading of the NT with the AFM cantilever. The dashed lines in (a), (b), and (d) represent fittings based on the core-shell model shown in Figure 3a.

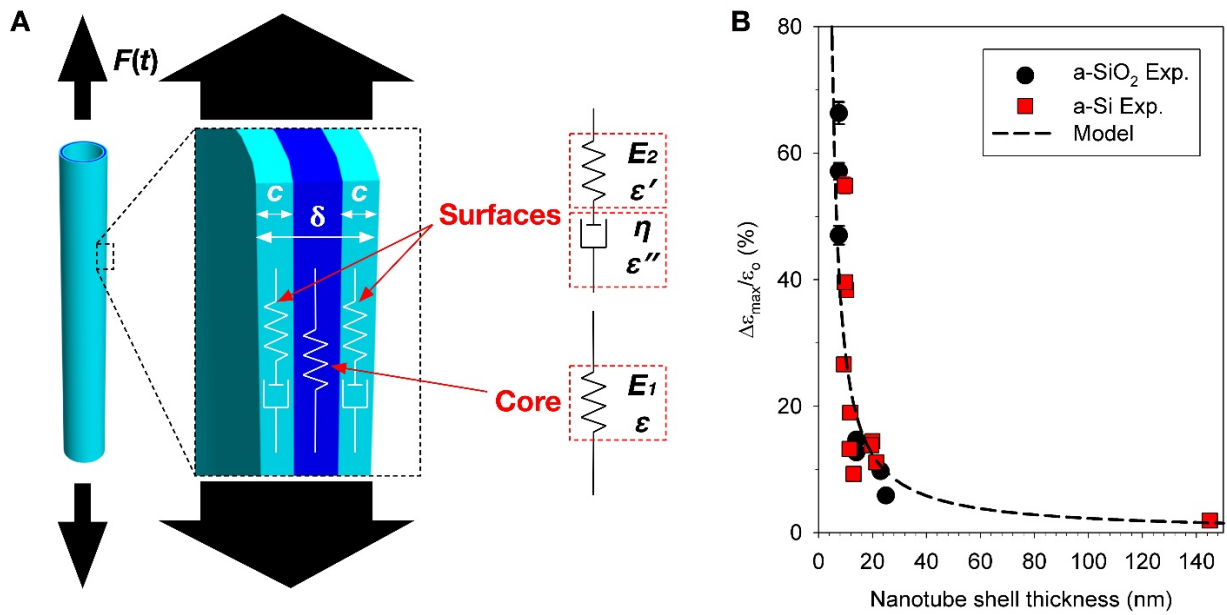

Figure 3: (a) Core-shell model describing the dynamic and size-dependent behavior of the amorphous nanotubes and (b) the experimental (circles and squares) and modeled (dashed line) size-dependent normalized saturated creep.

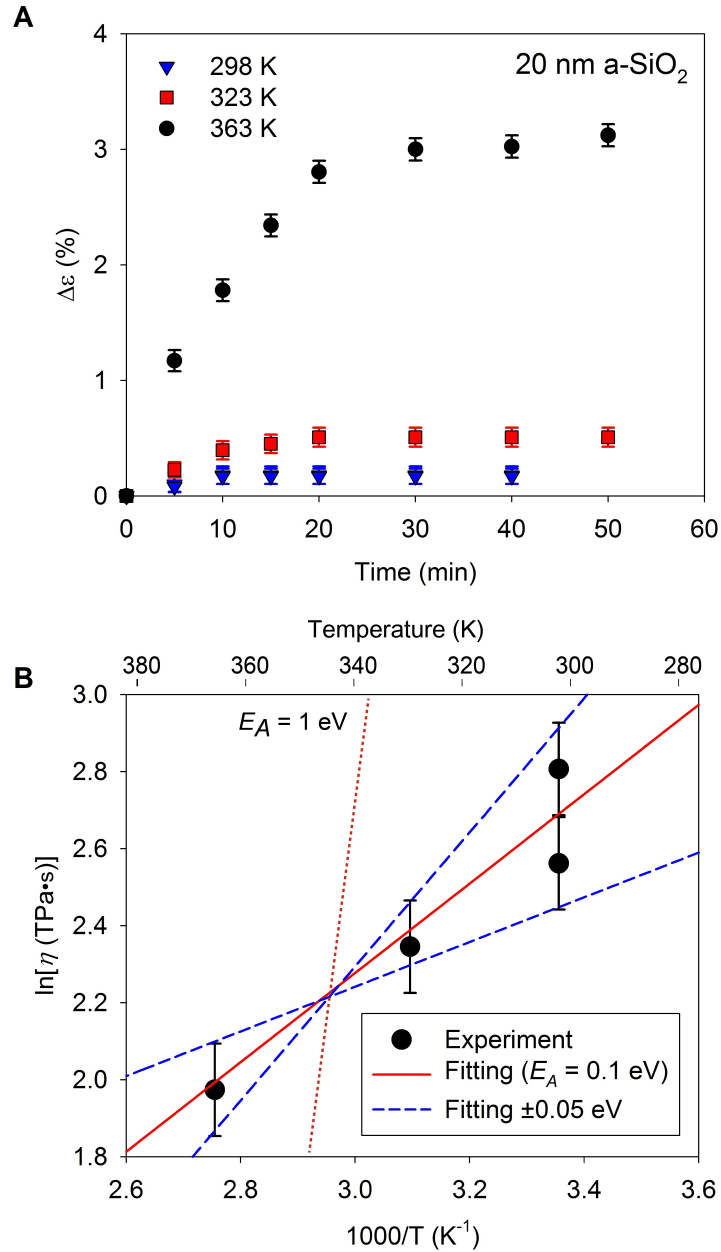

Figure 4: (a) Creep strain observed for 20 nm shell amorphous $SiO_2$ nanotubes at various temperatures. (b) Creep viscosity versus temperature, fitted to an Arrhenius temperature dependence with an activation energy of $E_A = 0.1 \pm 0.05$ eV, consistent with bond motion/rearrangement. For comparison, temperature dependent viscosity requiring an activation energy of 1 eV, signifying a bond breaking mechanism, is shown (short dashed line).